\documentclass{aa}
\usepackage{graphicx}
\usepackage{txfonts}
\begin{document}
   \title{Whiting~1:the youngest globular cluster associated
          with the Sgr dSph\thanks{Based on
          observations with the ESO VLT at the Paranal Observatory,
          under the program 76.D-0128.},\thanks{This paper includes
          data
gathered with the 6.5 meter Magellan Telescopes located at Las Campanas Observatory, Chile.}}

   \author{Giovanni Carraro
          \inst{1}
           \and
           Robert Zinn
           \inst{2}
           \and
           Christian Moni Bidin
           \inst{3}
          }

   \offprints{G. Carraro}

 \institute{
              Dipartimento di Astronomia, Universit\`a di Padova,
                  Vicolo Osservatorio 2, I-35122 Padova, Italy\\
                  \email{giovanni.carraro@unipd.it}
		 \and
              Astronomy Department, Yale University, P.O. Box
              208101, New Haven, CT 06511, USA\\
                  \email{robert.zinn@yale.edu}
 		  \and
              Departamento de Astronom\'ia, Universidad de Chile,
                 Casilla 36-D, Santiago, Chile\\
                  \email{mbidin@das.uchile.cl}
                }

   \date{Received ........; accepted....}

  \abstract
   {Recently, Carraro (2005) drew attention to the remarkable star
  cluster Whiting 1 by showing
that it lies about 40 kpc from the Sun and is therefore unquestionably
  a member of the Galactic halo (b=-60.6 deg.).
 Its Color Magnitude Diagram (CMD) indicated that Whiting 1 is very
  young ($\sim$5 Gyrs) for a globular cluster.
 It is very likely that Whiting 1 originated in a dwarf galaxy that
  has since
been disrupted by the Milky Way.
 }
   {The main goals of this investigation were to constrain better the
  age, metallicity, and distance of Whiting 1
and to assess whether it belongs to a stellar stream from the Sagittarius Dwarf Spheroidal Galaxy (Sgr dSph).
 }
  % methods heading )
   {Deep CCD photometry in the BVI pass-bands obtained with the VLT is
  used to improve the quality of the
CMD and to determine the cluster's luminosity function and surface density
  profile.  High-resolution spectrograms
obtained with Magellan are used to measure the cluster's radial velocity and to
  place limits on its possible metallicity.
  The
 measurements of distance and radial velocity are used to test the
  cluster's membership in the stellar
streams from the Sgr dSph.}
   {From our CMD of Whiting 1, we derive new estimates for the
  cluster's age ($6.5^{+1.0}_{-0.5}$ Gyrs),
 metallicity (Z=0.004$\pm$0.001, [Fe/H] = -0.65), and distance (29.4$^{+1.8}_{-2.0}$ kpc).
 From echelle spectrograms of three stars, we obtain -130.6$\pm$1.8 km/s for
  the cluster's radial velocity and show from measurements of two
  infra-red CaII lines that the [Fe/H] of the cluster probably lies in
  the range -1.1 to -0.4.
  Both the luminosity function and the surface density profile suggest
  that the cluster has undergone tidal stripping by the Milky Way. We
  demonstrate that the position of Whiting 1 on the sky, its distance
  from the Sun,
 and its radial velocity are identical to within the errors of both
  the theoretical predictions
 of the trailing stream of stars from the Sgr dSph galaxy and the
  previous observations of
the M giant stars that delineate the streams.}
   {With the addition of Whiting 1, there is now strong evidence that
  6 globular clusters formed within the Sgr dSph.
  Whiting 1 is
  particularly interesting because it
 is the youngest and among the most metal rich.  The relatively young
  age of Whiting 1 demonstrates
 that this dwarf galaxy was able to form star clusters for a period of
  at least 6 Gyr, and the
age and metallicity of Whiting 1 are consistent with the
  age-metallicity relationship in the
 main body of the Sgr dSph.  The presence now of Whiting 1 in the
  Galactic halo provides
 additional support for the view that the young halo clusters
  originated in dwarf galaxies
 that have been accreted by the Milky Way.}


   \maketitle

\section{Introduction}
The Milky Way galaxy harbors approximately 150 globular clusters
(Harris 1996 and later revisions)
 and thousands of open clusters.  Much of what we know about the
 evolution of the
Galactic stellar populations has been gained through the measurement
of the chemical compositions and the ages of these clusters.  We know,
for example, that the bulk of the globular clusters formed early in
the
 evolution of the Milky Way, roughly 12 Gyrs ago (e.g., Salaris \&
 Weiss 2002),
whereas the onset of star formation in the thin disk, as indicated by the
oldest
open clusters, started later, after a hiatus of a few Gyrs (Carraro et
al. 1999).
The ages of the metal-rich globular
clusters, some of which belong to the thick disk while others are
part of the bulge, fall roughly in this age gap.
However, there is a sizable spread in age among the globular clusters
in the halo,
and the
youngest of them, e.g., Palomar 1 (Rosenberg et al. 1998a), Palomar 12
(Rosenberg et al. 1998b), Ruprecht 106 (Buonanno et al. 1994)
and Whiting 1 (Carraro 2005), are significantly younger than the
globular cluster 47 Tuc, the prototypical thick disk globular cluster.
In fact, these young halo clusters
overlap in age with the oldest open
clusters.
  While the origin of these clusters is anomalous in a scenario where
the halo and the disk are the first and last stages of an evolutionary
sequence, they are explained naturally in the hierarchical
picture of galaxy evolution whereby large galaxies
are built through the accretion of many smaller ones (see Freeman \&
Bland-Hawthorne 2002 for a review).  The dwarf galaxies that were
accreted
experienced their own histories of star and star cluster formation.
 The galaxies accreted at the earliest epochs, the most numerous ones
 according to models (e.g., Bullock \& Johnston 2005), had a relatively
 short time to produce stars and clusters, whereas the ones accreted
 most recently had a larger span of time available.  It is therefore
attractive, as many other authors have, to identify the young globular
clusters as relatively recent additions to the Galactic halo that had
their origins in disrupted dwarf galaxies (e.g., Searle and Zinn
1978).
 The most clear-cut example of this is the current tidal destruction
 of the
 Sagittarius Dwarf Spheroidal (Sgr dSph) galaxy, which is probably the
 origin of Pal 12 (Dinescu et al. 2000; Martinez-Delgado et al 2002;
 Cohen 2004) and Whiting 1 (see below).
The recent detections of other stellar streams and substructures in
the
 halo, which appear to be unrelated to the Sgr dSph (Newberg et
 al. 2002, Duffau et al. 2006,
Belokurov et al. 2006,
Grillmair and Dionatos 2006; Grillmair 2006), is
 powerful
evidence that the destruction of Sgr is not an isolated event in an
otherwise
orderly evolution from halo to disk, but merely the most recent
episode of
 halo building through the accretion of satellite galaxies.

The youngest of the Galactic globular clusters, Whiting 1, was
discovered
recently by Whiting et al. (2002), who suspected it was a young open
cluster.
 More recently, Carraro (2005) demonstrated that this cluster cannot
 be in
 the Galactic disk, but must be in the halo.  From shallow photometry
 (V $\sim$ 22-23) obtained with the Yale 1.0m telescope at CTIO, he derived
 an age of roughly 5 Gyr, a metallicity of Z=0.004 ([Fe/H]= -0.7),
and a distance of $\sim$ 40 kpc from the Sun.  Since these are exceptional
 properties for an outer halo cluster, we have obtained deeper
 photometry
to constrain better these parameters.  We have also measured its
radial
 velocity so that its membership in the trailing stream of stars from
 the
Sgr dSph galaxy, which lies along the same line of sight, can be examined.

\begin{table}
 \fontsize{8} {10pt}\selectfont
 \tabcolsep 0.1truecm
 \caption{Radial velocity of the 3 Whiting~1 stars.}
 \begin{tabular}{cccccccc}
 \hline
 \multicolumn{1}{c}{ID}         &
 \multicolumn{1}{c}{$\alpha (2000.0)$} &
 \multicolumn{1}{c}{$\delta(2000.0)$}        &
 \multicolumn{1}{c}{$B$}       &
 \multicolumn{1}{c}{$V$}       &
 \multicolumn{1}{c}{$I$}       &
 \multicolumn{1}{c}{$\frac{S}{N}$} &
 \multicolumn{1}{c}{$RV$}       \\
 \hline
 & hh:mm:sec & $o$:$\prime$:$\prime\prime$& & & & & [km/sec]\\
\hline
  6& 02:02:57.1&  -03:15:07.7&     19.28&   18.00&  16.51& 30&  -131.1$\pm$2.8\\
  7& 02:02:52.7&  -03:15:56.2&     19.07&   18.15&  17.16& 30&  -130.5$\pm$3.5\\
  8& 02:02:55.2&  -03:14:48.7&     19.06&   18.14&  17.15& 30&  -130.1$\pm$3.1\\
 \hline
 \end{tabular}
\end{table}

%                                     Two column figure (place early!)
%______________________________________________ Gamma_1 (lg rho, lg e)
   \begin{figure}
   \centering
   \includegraphics[width=9cm]{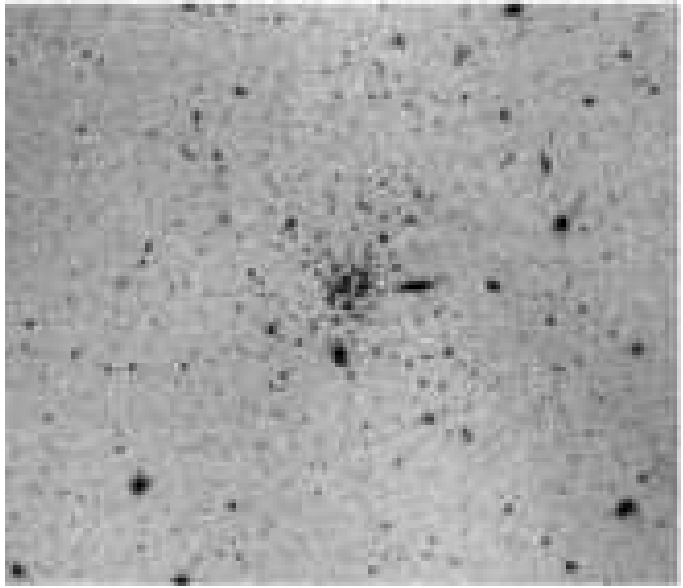}
   \caption{A  1200 secs B image of the Whiting 1 field. North is up,
    East to the left, and the image is 6.8 arcmin on a side.}
    \end{figure}

   \begin{figure}
   \centering
   \resizebox{\hsize}{!}{
   \includegraphics{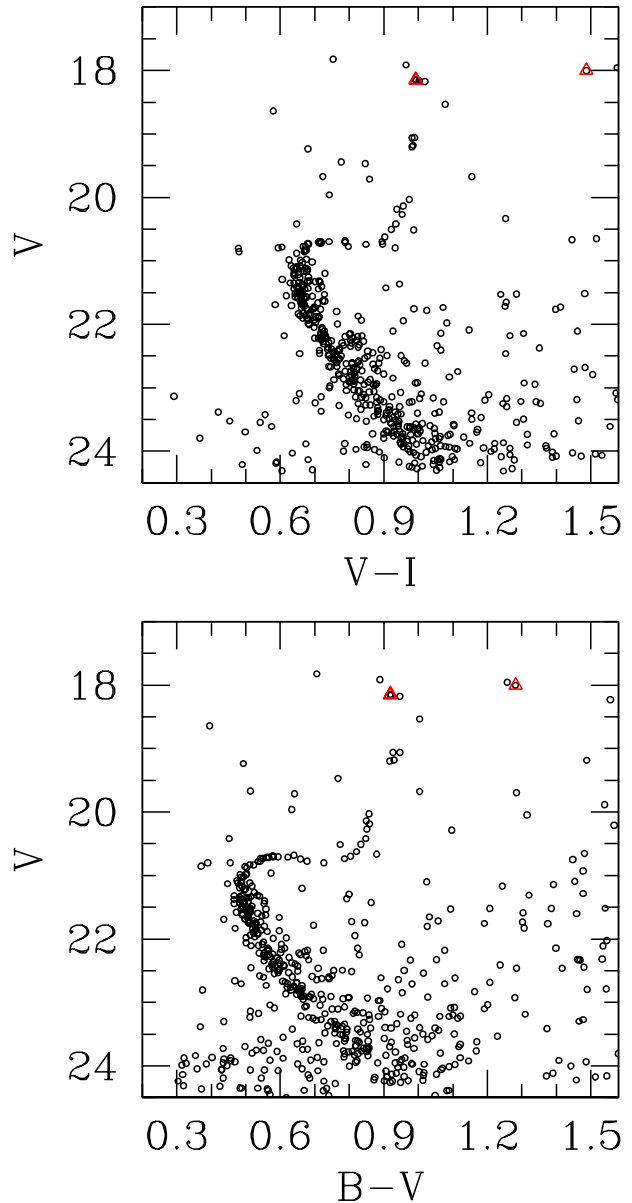}}
   \caption{CMDs in the V vs B-V (lower panel) and V vs (V-I) (upper
     panel) planes of the stars within a radius of 1 arcmin and having
     photometric errors lower than 0.06 in B, V and I. Open triangles indicate the 3
     stars observed spectroscopically.}
    \end{figure}

\section{Observations and Data Reduction}

\subsection{Photometry}

On the nights of 10 and 11 October 2005, Whiting 1 was observed in the
BVI filter bands with the Very Large
Telescope UT2 Kueyen and the FORS1 CCD camera. Six deep exposures were
taken (2x1200 sec in B, 2x700 sec in each
 V and I) on those nights, which had photometric conditions and an average seeing of 0.8 arcsec.  The camera has a scale
 of 0.2 arcsec per pixel and an array of 2048x2048 pixels.  An example
 of one of the 6.8x6.8 arcmin images
of Whiting 1 is shown in Fig. 1.  The standard IRAF \footnote{IRAF is distributed by the National Optical Astronomy Observatories, which are operated by the Association of Universities for Research in Astronomy, Inc., under cooperative agreement with the National Science Foundation.} routines were
used to reduce the raw images.
Using the psf-fitting routines of DAOPHOT and ALLSTAR (Stetson 1994) in
the IRAF environment, we
measured instrumental magnitudes for all of the stars in the field.
These magnitudes were
 transformed to the standard system using as standards the shallower, calibrated photometry of Carraro (2005).

\subsection{Color Magnitude Diagrams}

Carraro (2005) estimated that Whiting 1 has a radius of approximately
 1 arcmin.  The stars that
 we measured within that radius and having photometric errors less
 than 0.06 mag are plotted
in the CMDs in Fig.~2.  These diagrams are a substantial improvement over the ones measured by
Carraro (2005) in that they go roughly 4 mag. deeper and define more
 precisely the principal
sequences of the CMD.  The turn off (TO) from the main-sequence is
 quite clear and is located at
 V=21.2, (B-V)=0.48, and (V-I) = 0.63).  There appears to be a
 sequence running roughly parallel
 to the main-sequence that may be composed of binary stars.  Although
 the red giant branch (RGB)
 is sparsely populated, its lower part is well traced.  A clump of
 stars is located at V = 18.3,
 (B-V)=0.95, and (V-I) = 0.98, which is consistent with the core helium burning phase of evolution.

%______________________________________________ Gamma_1 (lg rho, lg e)
   \begin{figure}
   \centering
   \includegraphics[width=9cm]{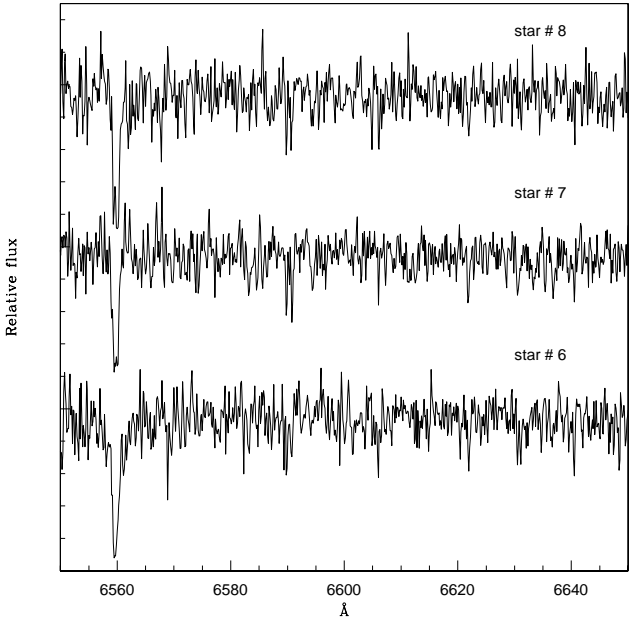}
   \caption{MIKE spectra of the 3 Whiting~1 stars (see also Table~1 and Fig.~2).}
    \end{figure}
%

%                                     Two column figure (place early!)
%______________________________________________ Gamma_1 (lg rho, lg e)

%

\subsection{Spectroscopy}

Echelle spectrograms of stars $\#$6, $\#$7, and $\#$8 (see Table 1 and Fig. 3)
 were obtained with
 integrations of 1200s per star on 2006 September 26 with the Magellan
 Inamori Kyocera Echelle (MIKE)
 spectrograph mounted on the Nasmyth focus of Landon Clay 6.5m
 telescope at the Magellan Observatory.
  Data were obtained with both the blue and red arms, but only the red
 spectra (4950-9500~ \AA)
had sufficient signal to noise ($\frac{S}{N}$) to be useful.  The spectra in
 part of this range are
 shown in Fig.3.  The slit was 1 arcsec wide, which yielded a resolution
 R=22000, and the CCD was
 binned in steps of 3 pixels in the dispersion direction.  The
 excellent seeing may have caused
 a higher resolution and a slight under sampling because of the
 binning.  We used quartz lamp
images with the diffuser in position for flat field correction, and
 the wavelength calibration
 was performed with ThAr lamp images that were taken just before and
 after three stellar exposures.
 The dark current was checked by examining several dark exposures and
 was found to be insignificant.
 The optimum algorithm (Horne 1986) was used to extract the spectra,
 which were also sky-subtracted
 and normalized using IRAF routines.

Radial velocity (RV) measurements were performed with the
 cross-correlation (CC) technique (Tonry \& Davis 1979)
 in the spectral range 5000-6700 \AA, which avoided the strong telluric
 bands at longer wavelengths.
We also made some measurements at other wavelengths and obtained
 results that were always consistent
 to within the errors.  For a template, we used a synthetic spectrum
 of a K3III star, but also checked
 the results with other spectra from the library of Coehlo et
 al. (2005) that differed in atmospheric
parameters and metallicity.  The height of the peak of the CC function
 decreased when there were
 relatively large differences between the template and target spectra,
 but the radial velocities remained stable.
  The results of the CCs were corrected for: 1) heliocentric
 correction, 2) systematic offsets in the wavelength
calibrations, which were estimated from measurements of the strong
 telluric bands in twilight sky exposures,
and shifts due to star centering in the slit, 3) a systematic offset
 ($\approx$1.0 km/s) from RV standard stars and from
 twilight sky exposures, which may result from the use of synthetic
 spectra in the CCs.  The errors quoted in
 Table 1 were calculated from the quadratic sum of all sources of
 error:  CC measurements,
wavelength calibrations, and the corrections stated above.  The RVs of
 the three stars are very similar
 to each other.  Under the reasonable assumption that all three are
 cluster members, we obtain from their mean
value a velocity of -130.6$\pm$1.8 km/s for the cluster.
   \begin{figure*}
   \centering
   \resizebox{\hsize}{!}{
   \includegraphics{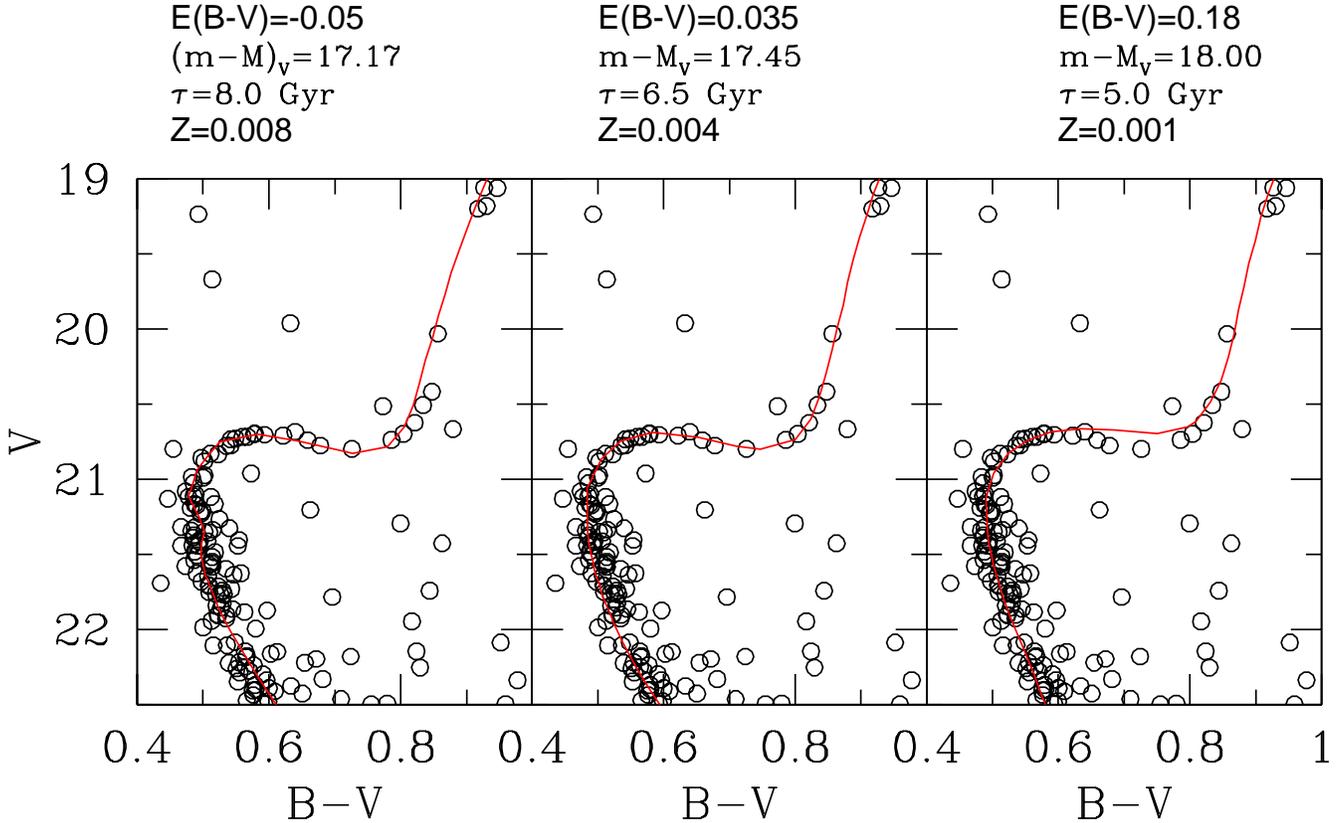}}
   \caption{Isochrone solutions for Whiting~1: three different
     age-metallicity combinations are explored, and each isochrone has
     been shifted by the reddening and apparent distance modulus
     indicated.}
    \end{figure*}

\section{Fundamental parameters}

\subsection{Estimates of reddening, age, metallicity, and distance modulus from the CMD}

By fitting isochrones for different ages and metallicity to the CMD of
Whiting 1, we can place limits on its
reddening and distance modulus.  In Fig. 4, the CMD of Whiting 1 is
compared with theoretical
isochrones that have been calculated by the Padova group (Girardi et
al. 20002) for Y=0.24 and
solar mixes of elements.  The metallicity was fixed, and then the age,
distance modulus,
 and reddening were adjusted until a match to the main-sequence,
 subgiant branch,
and lower RGB was obtained.  However, one of these fits produces
an unphysical value for the
 reddening, and another produces a very unlikely value.  The fit shown
 in the left panel is for a
 relatively high metallicity (Z=0.008, [Fe/H] = -0.37).  In order to
 match the relatively blue
colors observed for the Whiting 1 stars, it was necessary to assume
the cluster suffers from a
{\it negative} reddening.  The result obtained with a much lower metallicity
(Z=0.001, [Fe/H] = -1.30)
 is shown in the right panel of Fig.~4.  In this case, the Whiting 1
 stars are redder than the
 isochrones by a considerable amount, and to make them match up, it
 was necessary to assume a
 reddening of E(B-V) = 0.18.  This value is untenably larger than the
 value (0.03) that the dust
 maps of Schlegel et al. (1998) indicate is the average reddening in a
 6 sq. arcmin area around
 the cluster.  At the Galactic latitude of Whiting 1 (-60.64) only a
 small variation in
 reddening is expected over this area.  The results obtained with
 these two metallicities
 suggest that an intermediate value may produce reasonable results.
 The middle panel
 shows the results obtained with the metallicity Z=0.004 ([Fe/H] =
 -0.65), which produces
 a better match to CMD than either of the other two and also a
 reddening value that is
 consistent with the maps of Schlegel et al. (1998).

By fine tuning of the fit in the middle panel, we estimate that the
 age and metallicity of
 Whiting 1 is 6.5$^{+1.0}_{-0.5}$  Gyr. and Z=0.004$\pm$0.001, respectively.  By
 shifting the isochrone
 in the middle panel within acceptable limits, we estimate that the
 errors in E(B-V) and (m-M)$_V$
 are 0.01 and 0.1, respectively.  These values and an assumed ratio of
 total to selective (R=$\frac{A_V}{E(B-V)}$)
 of  3.1 yield 29.4$^{+1.8}_{-2.0}$ kpc for the heliocentric distance of
 Whiting 1.  If we place the
Sun at 8.0 kpc from the Galactic Center, Whiting 1 (l=161.62,
 b=-60.64) has a mean
Galactocentric distance of 33.9 kpc.
Its Cartesian coordinates (Harris 1996) are X=-21.7, Y=4.5, and Z=-25.6 kpc.

  \begin{figure}
   \centering
   \includegraphics[width=9cm]{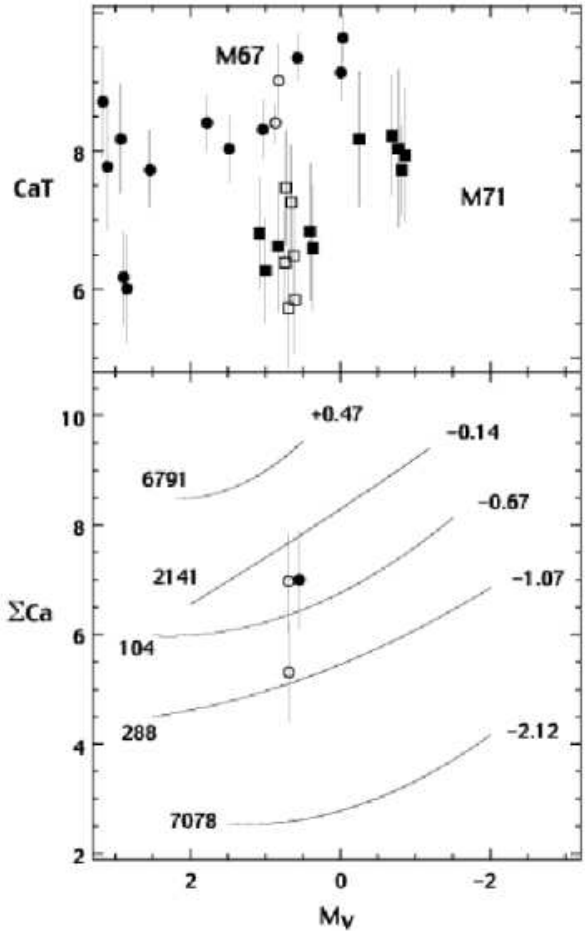}
   \caption{In the upper diagram, the CaT indices of Cenarro et al. (2001) for
red giants and red clump stars (solid and open symbols, respectively)
in M67 and M71 are plotted against Mv.  In the lower diagram, the $\Sigma$Ca
indices for the red giant and the red clump stars (solid and open
circles) in Whiting 1 are compared with the cluster sequences
observed by Carrera et al. (2007).  The [Fe/H] values that label the
cluster sequences are on the metallicity scale of Carretta \& Gratton
(1997).}
    \end{figure}

\subsection{Estimate of metallicity from the infrared CaII lines}

The $\frac{S}{N}$ of the spectrograms are too low for a determination of metal
abundance of Whiting 1 by
the standard technique of comparing the equivalent widths of weak
metal lines with synthetic spectra
calculations.  We can, however, determine a rough estimate from the
pseudo-equivalent widths of the
strong CaII lines in the near infrared, which were included in one of
the orders of the echelle spectrograms.
  Several studies have shown that measurements of these lines in
  low-dispersion spectrograms of
 red giants provide precise rankings of globular clusters by [Fe/H],
 and the recent studies by
 Pont et al. (2004), Cole et al. (2004), and Carrera et al. (2007)
 have shown that this method
can be applied without modification to red giants in open clusters that have ages greater than about 3 Gyr.

Only one ($\#$6) of the three stars observed in Whiting 1 lies near the
RGB.
The other two appear to be part of a red clump of core helium burning
stars.  Red clump stars
 (or red horizontal branch stars in globular clusters) have been seldom included in studies
 of the CaII lines, which have concentrated on stars on the RGB.
 However, a few of these stars
 in the open cluster M~67 and the globular cluster M~71 were included in
 the catalogue of CaII
 line strengths measured by Cenarro et al. (2001).  According to
 Sandquist (2004) and references
therein, M~67 is 4 Gyrs old and has approximately solar composition.
M71 is much older, $\sim$10 Gyrs,
 and more metal poor, [Fe/H] $\sim$-0.7 (e.g., Salaris \& Weiss 2002).  In
 the upper diagram of Fig. 5,
 we have plotted the index CaT measured by Cenarro et al. (2001)
 against M$_V$, using the distance
moduli listed by Sandquist (2004) and by Harris (1996).  This diagram
shows that the red clump
stars (or red HB stars) in M~67 and M~71 do not deviate systematically
from the sequence of RGB stars.
 While the RGB stars are cooler than the red clump stars of the same
 M$_V$, which weakens CaT in this
effective temperature range, they also have lower gravities, which
strengthens CaT (see Cenarro et al. 2002).
  The upper diagram in Fig.~5 suggests that the temperature and
  surface gravities differences offset each
 other to the precisions of the measurements.  There is therefore some
 justification for treating all
 three stars in Whiting 1 as if they belong to the RGB of the cluster.

One order of the spectrograms of the Whiting 1 stars cover the range
 8470-8710 \AA, and over this range
 the $\frac{S}{N}$ varies between $\sim$7 and $\sim$10.  While this low $\frac{S}{N}$ would preclude
 measurement of the CaII
 line strengths in the low-resolution spectrograms (R $\sim$ 2000) normally
 employed to measure them,
they are measurable in our much higher resolution spectrograms.  The
 wavelength ranges used by
 Cenarro et al. (2001) to define the CaT index extend both bluer and
 redder than the Whiting~1
spectrograms.  The ranges used by Armandroff \& Da Costa (1991,
 hereafter AD) to measure the two
 strongest of the CaII lines extend insignificant amounts beyond
 the limits of our spectrograms,
 and it is possible to measure their Ca II index.  However, AD did not
 observe any stars as low in
luminosity as the Whiting 1 ones.  Even lower luminosity ones have
 been measured recently by Carrera
et al (2007).  Although the CaII index ($\Sigma$Ca) defined by Carrera et
 al., which is based on the same
 side bands as the CaT index, is not measurable in our spectrograms,
 they have shown that there is a
tight correlation between the AD index and $\Sigma$Ca.  We have therefore
 followed closely the measuring
 technique of AD, which consists of fitting a straight line to the
 intensities in wavelength bands
on either side of the two strongest ($\lambda$8542 \& $\lambda$8662) of the three CaII
 lines.  The straight lines set
the pseudo-continua for the line profiles, which following AD were fit
 by Gaussian profiles.  To explore
 the effects of low $\frac{S}{N}$ on the measurements, we smoothed the
 spectrograms by varying amounts to mimic
 ones of higher $\frac{S}{N}$ but lower resolution and found no significant
 systematic offsets in the AD index.
 This is consistent with the results that Carrera et al. (2007) found
 when comparing spectrograms of
 the same stars taken with different spectrographs and resolutions.
 We also fit the Gaussian profiles
 to the Whiting~1 stars after varying the height of the
 pseudo-continua within reasonable limits to get
 an estimate of the uncertainties in our measurements. Finally, we
 used the transformation equation
 given by Carrera et al. (2007) to transform our measurements into values of $\Sigma$Ca.

In the lower panel of Figure~5, we compare the $\Sigma$Ca indices of the
Whiting~1 stars with the sequences
observed by Carrera et al. (2007) in 3 globular clusters (NGC 104,
288, 7078) and two open clusters
 (NGC 6791, 2141).  A distance modulus of (m-M)$_V$ = 17.45 (see above)
was adopted for Whiting~1.
All reasonable ones yield similar values of [Fe/H].  As expected from
the upper diagram, there
is no significant offset, to within the errors, between the two red
clump stars and the one RGB
 star in Whiting 1, and the mean position of the 3 stars suggests a
 metallicity near [Fe/H] = -0.7.
  The calibrations (Carrera et al. 2007) of $\Sigma$Ca in terms of [Fe/H] on
  the metallicity scales of
 Carretta \& Gratton (1997), Kraft \& Ivans (2003), and Zinn \& West
 (1984) yields mean values of
-0.65$\pm$0.20, -0.82$\pm$0.20, and -0.83$\pm$0.26, respectively.  These errors
include our estimates of
the random errors in our measurements and the transformation of the AD
indices to $\Sigma$Ca.
We caution that they may be too small because we do not have any
repeated observations
or observations in common with other observers that would enable us to
measure better
our random errors and correct for any systematic ones.  Nonetheless,
these measurements
indicate that the [Fe/H] of Whiting 1 probably lies within the range
of -0.4 to -1.1,
 which is consistent with what we found above from comparing isochrones to the CMD.

\begin{table}
 \fontsize{8} {10pt}\selectfont
 \tabcolsep 0.1truecm
 \caption{Completeness results for three different zones
in Whiting 1.}
 \begin{tabular}{cccc}
 \hline
 \multicolumn{1}{c}{$V$}         &
 \multicolumn{1}{c}{$R \leq 0^{\prime}.75$}       &
 \multicolumn{1}{c}{$0^{\prime}.75 \leq R \leq 1^{\prime}.50$} &
 \multicolumn{1}{c}{$1^{\prime}.50 \leq R \leq 2^{\prime}.25$}       \\
 \hline
21.0&    95.2&   99.9&   99.9\\
21.5&    96.9&   99.9&   99.8\\
22.0&    93.6&   98.6&   98.2\\
22.5&    95.0&   97.5&   97.1\\
23.0&    88.5&   88.1&   96.5\\
23.5&    79.6&   82.1&   92.4\\
24.0&    62.5&   70.9&   89.1\\
 \hline
 \end{tabular}
\end{table}

\subsection{The Luminosity Function}

With the goal of measuring the integrated apparent magnitude of
Whiting 1 (V$_t$) and hence its absolute magnitude,
we have determined its luminosity function (LF) in different radial
zones.  To do this, we estimated the
completeness of our photometry by running experiments with artificial
stars.  First, we divided the cluster into 3
concentric rings centered on the cluster:  r $\leq$ 0.75,
$0.75\leq r \leq1.50$, and $1.50\leq r \leq2.25$.
Within these rings, we counted 204, 127, and 107 stars brighter than V
=24.  We then used the routine ADDSTAR
within DAOPHOT to insert 1000 artificial stars at random positions
over the whole field and
within the magnitude range of the real stars.  This number of
artificial stars was chosen so that
 a reasonable number of them (about 30\% of the real star population)
 would lie within the cluster radius
($\sim$1.5 arcmin.).  This was done for both the long and the short
 exposures.  We then reduced the images
 with the artificial stars in exactly the same manner as the real
 images.  The ratio of the number of
artificial stars recovered by ALLSTAR to the number inserted defines
the completeness.  This experiment
 was run 10 times using a different seed number with the random number
 generator that ADDSTAR uses
in the calculation of the positions and magnitudes of the artificial
stars.  The mean values that
 were found for the completeness in the three radial zones are listed in Table~2.

To construct the faint part of the LF, we placed two curves on either
side of the subgiant branch
and the main-sequence and parallel to the ridge line defined by the
concentration of cluster stars.
We then counted the number stars within this band that have $21.0\leq V \leq24.0$.
In exactly the same way,
we counted the stars in the field, far beyond the cluster radius, that
lie in the same region of the CMD.
This field contribution was normalized to the areas of the radial
zones and subtracted from
each of the magnitude bins of the LF.  The background and completeness
corrected LF for the three
radial zones is shown in Fig. 5, where the error bars are the ones
indicated by Poisson statistics.
In the innermost and outermost regions and nearly so in the
intermediate region, the LF
flattens out toward faint magnitudes.  In {\it normal} globular clusters
the LF continues to rise.  Flat LFs that resemble Whiting 1's have been observed in globular clusters that
appear to be undergoing
tidal stripping by the gravitational field of the Milky Way (e.g., Pal 5, Koch et al. 2004; Pal 13, Cote et al. 2002).

To obtain V$_t$, we combined the three LFs in Fig. 5 and added the
brighter stars that lie either close to expected
location of the RGB or are part of the red clump.  Integration of this
combined LF yields Vt = 15.03.
This value and the distance modulus of (m-M)$_V$ = 17.45 yield M$_V$ =
-2.42 for Whiting 1.
The uncertainty in this value is $\pm$0.1 and possibly larger, for we
cannot be certain that all
of the bright stars have been included or that the field stars have
been removed.
Despite this uncertainty, it is clear that Whiting~1 is among the
least luminous globular clusters known.
Among the $\sim$150 Galactic globular clusters,
only 4 (Pal 1, E3, AM4, Willman 1, Segue 1) in addition to Whiting 1
have
M$_V$ larger than -3 (Harris 1996, 2003 ed.; Willman et al. 2006; Belokourov et al. 2006).

\subsection{The surface density profile}

If, as suggested by its LF, Whiting 1 has experienced tidal stripping,
its surface density profile is
expected to be also abnormal.  This profile was constructed by
counting the stars in our catalogue of V
photometry that lie within the color range $0.6 \leq V-I \leq1.4$ and are
brighter than V=24.0.
The paucity of cluster stars (see Fig. 1) made uncertain the location
of the cluster center,
and we defined it by the peaks in the histograms of the number of star
images per distance interval in the
X (RA) and Y (DEC) directions.  To measure the background of field
stars, we subtracted from the
catalogue the stars lying less than 1.5 arcmin from the cluster
center (a preliminary radius) and
then counted the remaining stars that fell within the same magnitude
and color ranges that were
used when counting the cluster stars.  This yielded 3.35 stars per
square arcmin for the background,
which is indicated in Fig. 7 by the symbol $\Sigma_{B}$.  Contributing to this
background are stars belonging
to the trailing stream of stars from the Sgr
dSph galaxy, in which Whiting 1 is immersed (see below).\\
\noindent
Unlike the surface density profiles of many globular clusters, the
profile of Whiting 1 (Fig. 7) cannot be approximated with a single
King profile.
If a King profile is fit to the outer part (log R $\geq$ 1.8 arcsec),
it grossly
underestimates the density in the inner part, which is more secure
statistically.
We show in Fig. 7 the opposite case where a King profile is fit to the
innermost 4 points,
which produces a very poor fit to the outermost points.  Compared to
this profile there
is an excess of stars that follow a power law (R$^{-\gamma}$) density
profile with $\gamma \approx$ 1.8.  Other globular clusters that are
suspected to have undergone
tidal stripping (see Grillmair et al. 1995; Cote et al. 2002) have
profiles that are
similar to the one of Whiting 1.  Their inner parts are well
represented by a King profile,
but beyond the tidal radius of the profile there are {\it
  extra-tidal} stars that
follow a power law slope.  Since this behavior is also expected
theoretically
(e.g., Johnston et al. 1999), we conclude that the surface density
profile of Whiting 1,
as well as its LF, suggests that it is losing mass by tidal stripping.

   \begin{figure}
   \centering
   \includegraphics[width=9cm]{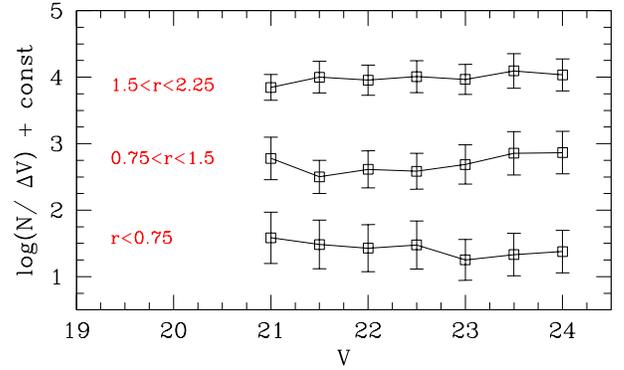}
   \caption{Completeness and background corrected LF of Whiting~1
   for the three indicated rings.  For clarity, the LF's for the outer and intermediate rings have been shifted upward by 3.0 and 1.5, respectively.}
    \end{figure}

   \begin{figure}
   \centering
   \includegraphics[width=9cm]{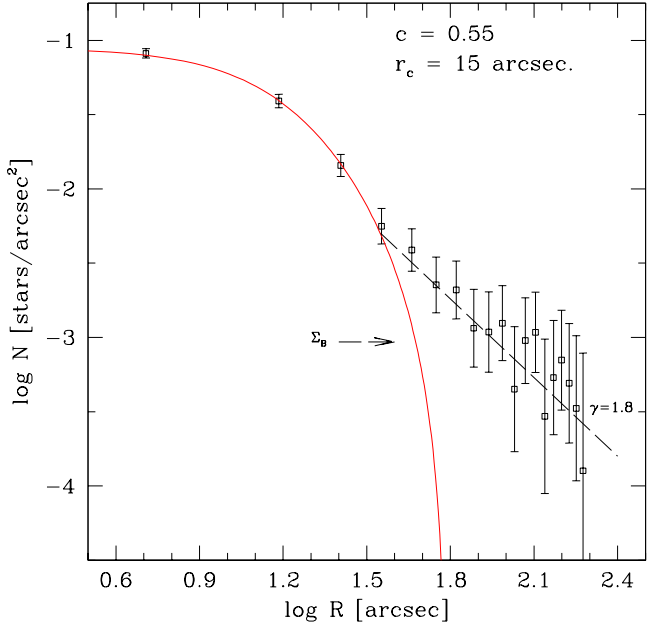}
   \caption{Radial surface density profile of Whiting~1. The solid
line is a King profile, while the dashed line is a power-law profile
for the extra-tidal stars. The density of the background is indicated by
$\Sigma_B$. See text for more details.}
    \end{figure}

\section{Membership in the Sagittarius Trailing Stream}
As discussed by Carraro (2005) and above, Whiting 1 is several Gyrs
younger than any other star cluster in the Galactic halo.  While the
presence of such a young object in the halo would have once seemed
mysterious, this is no longer so because the halo is now known to be
crisscrossed with stellar streams from disrupted satellite galaxies
(see above).  We show below that the position and radial velocity of
Whiting 1 are consistent with membership in the stream of stars
trailing behind the Sgr dSph galaxy as it orbits the Milky Way.\\
\noindent
In order to compare Whiting 1 with models of the Sgr streams and with
stars that constitute the streams, we have computed its longitude
($\lambda_{\odot}$) and latitude ($\beta_{\odot}$) in the Sgr orbital
plane as viewed from the Sun, and its radial velocity (Vgsr) in the
Galactic rest frame ($\lambda_{\odot}$ =102$^o$.91,
$\beta_{\odot}$=+1$^o$.33, Vgsr = -105.0 km sec-1).  In these
calculations, we have followed the procedures and the choice of solar
motion used by Majewski et al. (2003, 2004)\footnote{See also
http://www.astro.virgina.edu/~srm4n/Sgr} so that the results for
Whiting 1 could be directly compared with theirs for M giants and with
the theoretical models of Law et al. (2005).  In the Sgr coordinate
system, the main body of Sgr is located at $\lambda_{\odot}$= 0$^o$, and
the $\lambda_{\odot}$ increases in the direction of the trailing
stream.  The coordinates for Whiting 1 show that on the sky it lies in
the direction to the trailing stream and very close to the orbital
plane.\\

\noindent
In upper panel of Fig.~8 is a plot of distance from the Sun (d) against
 $\lambda_{\odot}$ in which we have plotted Whiting 1, M giants with measured
radial velocities (Majewski et al. 2004), and the model that Law et
al. (2005) calculated under the assumption that the gravitational
potential of the Milky Way is spherical in shape.
The lower panel compares the values of Vgsr for these objects
with the same models.  Law et al. (2005) have also calculated models
assuming prolate and oblate potentials.  Over the ranges of
 $\lambda_{\odot}$  and
$d$ that are relevant to Whiting 1 these models
produce similar distributions in the diagrams
plotted in Fig.~8 (see figs. 2 and 10 in Law et al. 2005).  The models of
Law et al (2005) are based in part on distribution of the M giants,
but over much larger regions of the sky than
are included in Fig.~8.  Therefore, the less than perfect match
of the model to the M giants in this restricted region is not
surprising.
This disagreement could also be due to an underestimate of
the distances to these stars (see Chou et al. 2006).\\
\noindent
The data plotted in Fig.~8 indicate that not only does Whiting 1 lie
in the direction to the Sgr trailing stream, but its distance places
it within the stream of M giants and its radial velocity suggests that
it and the M giants are moving on similar orbits.  These data for
Whiting 1 agree almost perfectly with the predictions of the models
constructed by Law et al. (2005) for the Sgr streams (see Fig. 8), and are
also consistent with the models calculated by Martinez-Delgado et
al. (2004a) and by Helmi (2004a,b).
On the
basis of the above evidence, we believe that it is highly probable
that Whiting 1 originated in Sgr, and in the following section, we
compare its properties with the other Sgr clusters.

   \begin{figure}
   \centering
   \includegraphics[width=\columnwidth]{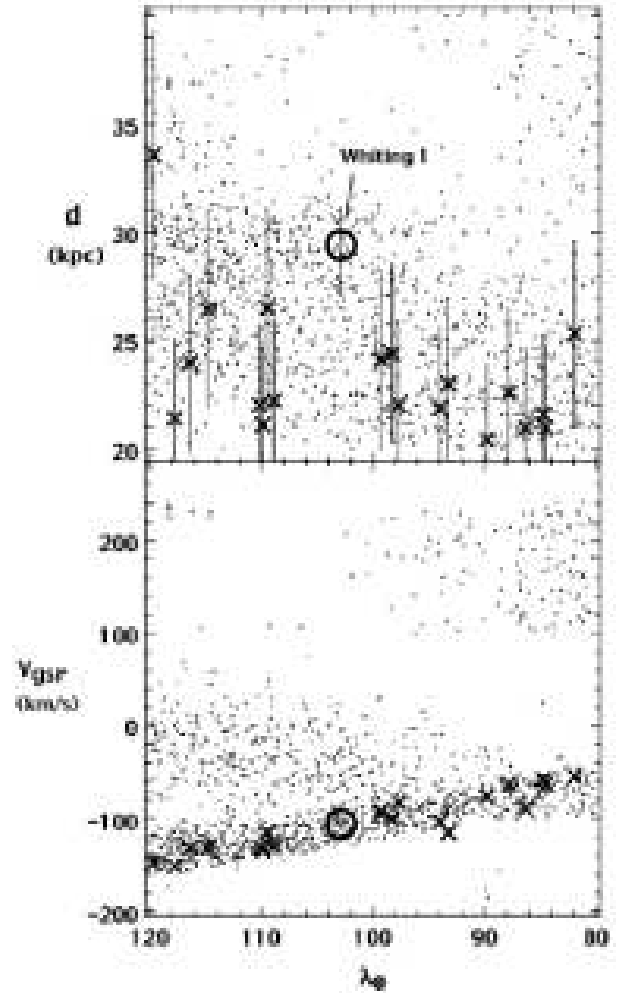}
   \caption{Whiting 1 (open circle) and M giants (X's) from Majewski et al.
(2003,2004) are compared with the model of the Sgr stream (small points) that
was computed by Law et al. (2005) under the assumption that the Galactic
potential is spherical.  For clarity, only objects between 20 and 40 kpc
from the Sun have been plotted in the two diagrams.  Distance from the Sun
(d) and velocity in the Galactic rest frame (Vgsr), are plotted against
longitude in the Sgr orbital plane ($\lambda_{\odot}$ in deg.).  The error bars of
the velocity measurements have not been plotted because they are smaller
than the symbols.}
    \end{figure}

   \begin{figure}
   \centering
   \includegraphics[width=9cm]{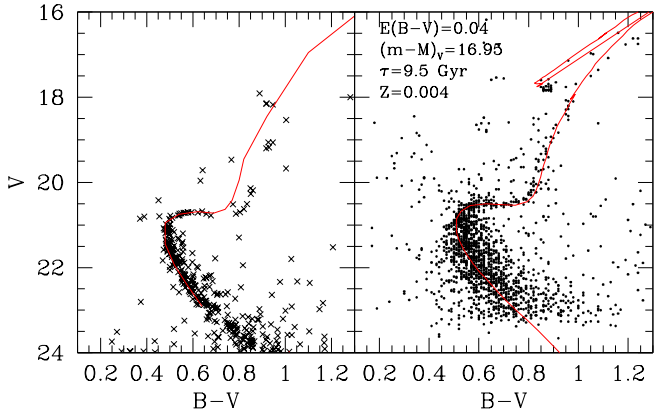}
   \caption{{\bf Left panel}: Whiting~1 stars with the fiducial sequence (solid line)
   of Terzan~7 (Buonanno et al. 1995) superimposed.; {\bf
   Right panel}: Terzan~7 stars with a Z=0.004 isochrone
   (solid line) for the age of 9.5 Gyr superimposed. The isochrone has been shifted by the
   values of E(B-V) and (m-M)$_V$ indicated}
    \end{figure}

\section{Whiting 1 and the star cluster family of the Sgr dSph}
The main body of the Sgr dSph contains 4 globular clusters:  M54,
  Terzan 7 and 8, and Arp 2.
  M54, which lies at the center of the main body, is very luminous and
  has an age similar
 to the ages of the oldest globular clusters in the Galactic halo.  It
  is unusual for
 a globular cluster in having an internal metallicity spread, and
  several authors
 (e.g., Layden and Sarajedini 2000) have suggested that M54 is
  actually the nucleus
 of the galaxy, which if correct would make Sgr a nucleated dwarf
  elliptical galaxy.
  Two of the other star clusters, Ter 8 and Arp 2, are metal-poor and
  old globular
 clusters, whereas Ter 7 is both metal-rich and young for a globular
  cluster
 (Buonanno et al. 1995).

Several authors have examined the sample of Galactic globular clusters
 for ones that may have originated in the Sgr dSph galaxy, with in
 some
 cases conflicting conclusions (e.g., Palma et al. 2002,
 Bellazzini et al. 2003; Martinez-Delgado et al. 2004b).
The young and metal-rich Galactic globular cluster Palomar 12 appears
to be the most likely case of a detached Sgr cluster, although some
others cannot be ruled out with certainty.  Dinescu et al. (2000) have
shown that the orbit of Pal 12 is consistent with it once being a
 member of Sgr.
  The photometry by Martinez-Delgado et al. (2002) indicates that Pal
 12 is
 embedded in a rich population of extra-cluster stars at the same
 distance
 from the Sun as the cluster, which is a sign that the cluster is part
of stellar stream.  The high-dispersion spectroscopy of red giants in
 Pal 12 by Cohen (2004) indicates that it has a chemical composition
 that
is unusual for a Galactic globular cluster, but perfectly compatible
 with
 an origin in the Sgr dSph galaxy

Since Ter 7 and Pal 12 are the youngest and most metal rich of the Sgr
 family of star clusters, including the other clusters suspected to be
 members,
 they are the ones that are most interesting to compare with Whiting
 1.
 While there has been no disagreement that both Ter 7 and Pal 12 are
 young
 and metal rich, the ages and metallicities that have been quoted for
 these
 clusters in the literature do vary considerably.  Their unusual
 chemical
 compositions may explain why the application of the methods that are
 used
 to rank the metallicities of Galactic globular clusters from the
 properties
 of the red giant branch and from low dispersion spectroscopy have
 yielded
 conflicting results.  The high dispersion spectroscopy of red giants
 in
 Ter 7 by Sbordone et al. (2005) and by Tautvaisiene et al. (2004)
have yielded [Fe/H] = -0.6.  Spectroscopy of red giants in Pal 12
by Cohen (2004) indicate that it has [Fe/H] = -0.8.  According to
these groups, both clusters have approximately solar $\alpha$/Fe, which
is consistent with field stars in the main body of Sgr that have
 similar
[Fe/H] (Bonifacio et al. 2004; McWilliam and Smecker-Hane 2005; Monaco
 et al. 2005).
 Since we have estimated above that Whiting 1 has essentially the same
metallicity as Ter 7, which is not far from that of Pal 12, it is
 relatively straightforward to compare the ages of these three
 clusters by simply overlaying of their CMD's.
We have done this first for Pal 12 and Ter 7 by selecting the CMD's of
 Stetson et al. (1989) and Buonanno et al. (1995), respectively.
By shifting the ridge-lines for Pal 12 by +0.07 in (B-V) and +0.6 in
 V,
 the TO regions and the horizontal branches (HB's) of Pal 12 and
Ter 7 are brought into coincidence.  With these shifts, there are only
very small differences in magnitude and color between the ridge-lines
of Pal 12 and Ter 7.   The lower red giant branch of Ter 7 may lie
slightly to the red of Pal 12's, which is consistent with Ter 7 being
 more metal-rich by a small amount, as suggested by the spectroscopic
 studies.
  However, this could also be nothing more than observational error.
 The overall excellent match of ridge-lines indicates that Ter 7 and
 Pal 12 are
 essentially coeval ($\Delta$t less than 1 Gyr).

The CMD of Whiting 1 is compared with the ridge-line of Ter 7
 (Buonanno et al. 1995)
 in the left panel of Fig. 9.  Because the HB of Whiting 1 is defined
 by very few
stars and resembles more a red clump than the HB of a globular
 cluster, we have
 not used it in this comparison.  Instead, the ridge-line of Ter 7 was
shifted by -0.04 in B-V and +0.2 in V in order to bring the TO
regions of the two clusters into coincidence.  Fig. 9 shows that after
 registration of the TO regions, there is a large difference
 between
 the red giant branches of Whiting 1 and Ter 7 in the sense expected
 if
Whiting 1 is the younger cluster.  How much younger is estimated in
 the
 right-hand panel of Fig. 9, where an isochrone from the same set that
was used to date Whiting 1is compared with the photometry of Ter 7 by
Buonanno et al. (1995).  By fine-tuning the fit, we estimate the age
 of Ter 7 is 9.5$^{+0.5}_{-1.0}$ Gyrs., which suggests that Whiting 1 is
 approximately 3 Gyrs. younger than Ter 7 and Pal 12,
 which as noted above are near clones.

The metal-poor clusters in Sgr appear to be approximately coeval with
 the oldest
 Galactic globular clusters (e.g., Layden and Sarajedini 2000), which
 according
 to recent studies (e.g., Salaris and Weiss 2002) have ages of 12 to
 13 Gyrs.
 The identification of Whiting 1 with Sgr suggests that this galaxy
 was able to
 form star clusters for  $\approx$ 6 Gyrs.  The ages and metallicities of Ter 7
 and
Pal 12 indicate that [Fe/H] rose relatively rapidly in over the first
 few
 Gyrs to $\approx$ -0.6.  It appears to have been essentially level from the
 formation
 of those clusters until at least the formation of Whiting 1 $\approx$ 3
 Gyrs. later.
 This is consistent with the recent conclusion of Bellazzini et
 al. (2006)
that the stellar population in the core of Sgr has [Fe/H] $\approx$ -0.6 and
 age $\approx$ 8
 Gyrs.  Observations of the red giants clearly indicate that [Fe/H]
 eventually rose to a higher value ($\approx$ 0) than found in any of the known Sgr
 clusters
(e.g., Bonifacio et al. 2004; McWilliam and Smecker-Hane 2005; Monaco
 et al. 2005).
Because there is a gradient of decreasing [Fe/H] with distance from
 the center of
the main body, and because objects in the streams are most likely
 detached from
the outskirts of the galaxy (see Chou et al. 2006), the use of Whiting
 1 and Pal
 12 to anchor the age-[Fe/H] relation must be viewed with some caution.

The on-going destruction of the Sgr dSph has so far deposited in the
Galactic
halo two remarkably young globular clusters, Pal 12 and Whiting 1.
Its
final destruction will contribute another young cluster Ter 7, the
more
 normal ones Arp 2 and Ter 8, and M54, the suspected nucleus.  It is
 possible
that other Sgr clusters remain to be discovered in its tidal streams
and
 that some Galactic globular clusters besides Pal 12 and Whiting 1
will be definitely identified with Sgr.  The drama being played out
today
by the destruction of this dwarf galaxy may be characteristic of many
earlier
 accretion events, which are speculated to have populated the Galactic
 halo
 with stars and star clusters (e.g., Bullock and Johnston 2005; Searle
 and Zinn 1978).
 Since several other stellar streams have been already identified in
 the halo,
the identification of star clusters with these or other streams may not be far behind.

\section{Summary and Conclusions}
The photometry reported here has confirmed the major conclusions of
 Carraro (2005) that Whiting 1
 lies in the outer Galactic halo and is unusually young for a globular
 cluster.
We derived here a somewhat larger age for the cluster (6.5 Gyrs) and a
 smaller distance from the Sun (29.4 kpc).
 The luminosity function and the surface density profile that we
 obtained suggest that Whiting 1 has
 been losing mass via tidal stripping by the Milky Way.  The new
 distance to Whiting 1 places
 it within the trailing stream of stars from the Sgr dSph galaxy.
The radial velocity that we have measured for the cluster is very
similar to the ones of M giants in the stream
and is in excellent agreement with  theoretical predictions for the
stream at its location.
The age and metallicity derived here for Whiting 1 are consistent with
the age-metallicity relationship of Sgr.
We conclude that Whiting 1 is another example of a young globular
cluster that has been detached from
its parent dwarf galaxy and is now a member of the Galactic halo.

\begin{acknowledgements}
This research was part of a joint project between Universidad de
Chile and Yale University, partially funded by the Fundacion Andes.
RZ was supported by National Science Foundation grant AST-05-07364.
GC was partially supported by Fundaci\'on Andes.
CMB acknowledges University of Chile graduate fellowship support from
programs MECE Educaci\'on Superior UCH0118 and Fundaci\'on Andes C-13798.
\end{acknowledgements}

\end{document}